\documentclass[sigconf]{acmart}

\AtBeginDocument{%
  \providecommand\BibTeX{{%
    \normalfont B\kern-0.5em{\scshape i\kern-0.25em b}\kern-0.8em\TeX}}}

\setcopyright{rightsretained}
\copyrightyear{2018}
\acmYear{2018}
\acmDOI{}

\acmConference[TCC BSI '18]{Trabalho de Conclusão de Curso (Graduação em Sistemas de Informação)}{Dezembro, 2018}{Recife, PE}
\acmBooktitle{Trabalho de Conclusão de Curso (Graduação em Sistemas de Informação),
  Dezembro, 2018, Recife, PE}
\acmPrice{}
\acmISBN{}

\usepackage[utf8]{inputenc}
\usepackage[brazil]{babel}

\begin{document}

\title{Integração e Entrega Contínua para aplicações móveis desenvolvidas em React Native}

\author{Pedro José de Souza Neto}
\affiliation{%
  \institution{Centro de Informática, Universidade federal de Pernambuco}
  \city{Recife}
  \state{Pernambuco}
  \country{Brasil}
}
\email{pjsn@cin.ufpe.br}

\author{Vinicius Cardoso Garcia}
\affiliation{%
  \institution{Centro de Informática, Universidade federal de Pernambuco}
  \city{Recife}
  \state{Pernambuco}
  \country{Brasil}
}
\email{vcg@cin.ufpe.br}

\renewcommand{\shortauthors}{Souza Neto e Garcia}

\begin{abstract}
  Continuous integration and continuous delivery are not new for developers who create web applications, however in the development of mobile applications this practice is still not very common mainly because of the challenges during the process of distributing the application. In the face of the growing number of applications, a greater requirement for quality and ever-shorter delivery times, delivering a healthy code is often extremely important to keep up with the competition. The purpose of this work is to implement an integration and continuous delivery pipeline for mobile applications developed in React Native. It intends to automate the process of build and delivery of applications developed with this technology.
\end{abstract}



\maketitle
\section{Introdução}

O crescente número de aplicações de software, uma exigência maior em relação a qualidade e com os prazos de entrega cada vez menores, acarretou em uma mudança que afetou o processo de desenvolvimento e a maneira que o software é entregue para o cliente.
Diante desse cenário, vimos o surgimento de metodologias como DevOps \cite{Ebert2016}, que tem como base a prática de integração e entrega contínua no seu ciclo operacional com a finalidade de aumentar a capacidade de resposta às necessidades dos clientes por meio de lançamentos de software frequentes e automatizados [10].
Com a propagação dessas práticas supracitadas, muitas empresas desenvolveram diversas soluções aplicando integração contínua \cite{Duvall2007, Zhao2017} e entrega contínua \cite{Chen2017} para aplicações web, porém no mundo de aplicações móveis, a adoção dessas práticas ainda é pouco utilizada \cite{Klepper2015, Jacksman2020}. 

O objetivo deste trabalho é o estudo e a implementação de uma pipeline de integração e entrega contínua para aplicações móveis desenvolvidas em React Native \cite{ReactNative} com fins de automatizar o processo de \textit{build} e entrega destas aplicações tendo os seguintes objetivos específicos:
\begin{itemize}
    \item Criação da pipeline da aplicação
    \item Estudo de ferramenta de integração e entrega contínua
    \item Estudo de ferramenta para testar aplicações React Native
    \item Implementação de uma aplicação em React Native e dos testes baseados nos estudos para realizar as funções desejadas
\end{itemize}

Para o desenvolvimento da \textit{pipeline}, foram utilizadas as ferramentas Jenkins\footnote{\url{https://www.jenkins.io/}, último acesso em 29/03/2021}, que é o principal servidor de automação \textit{open source} que fornece centenas de \textit{plugins} para apoiar os processos de \textit{build}, implantação e automação para diversos tipos de projeto; Blue Ocean\footnote{\url{https://www.jenkins.io/projects/blueocean/}, último acesso em 29/03/2021}, que repensa a experiência do usuário do Jenkins, reduz a desordem e aumenta a clareza para todos os membros da equipe; TestFairy\footnote{\url{https://www.testfairy.com/}, último acesso em 29/03/2021}, que é uma plataforma de testes para aplicações móveis que fornece registros e relatórios de falhas além de ser um meio de centralizar a distribuição do aplicativo; e, finalmente, o Slack\footnote{\url{https://slack.com/}, último acesso em 29/03/2021}, que é uma canal de comunicação presente em quase todas as empresas além de possuir ótimos meios de integração com as demais ferramentas utilizadas nesse trabalho. 

Esse documento está organizado como se segue: Seção 1 - Introdução: Apresenta motivações e objetivos do estudo; Seção 2 - Conceitos básicos: Apresenta um contexto teórico e tecnológico para a compreensão dos capítulos seguintes; Seção 3 - Implementação do aplicativo: Apresenta a escolha da tecnologia adotada, o desenvolvimento dos testes e do aplicativo; Seção 4 - A pipeline de integração e entrega contínua: Apresenta o passo a passo no processo de construção da pipeline; e, finalmente, Seção 5 - Conclusões: Apresenta as conclusões obtidas durante o desenvolvimento do projeto, trabalhos futuros e limitações.

\section{Conceitos Básicos}
Nesta seção são apresentados e discutidos os conceitos básicos que fundamentam este trabalho.

\subsection{Frameworks de desenvolvimento em JavaScript}
\label{subsec:frameworks}

No princípio, o JavaScript\footnote{\url{https://en.wikipedia.org/wiki/JavaScript}, último acesso em 29/03/2021} era utilizado na construção de animações e para fornecer interatividade para algumas páginas da web. Contudo, não existia nenhum tipo de comunicação com o servidor, todas as ações eram voltadas apenas para exibição de artefatos.

Com o advento do AJAX\footnote{\url{https://en.wikipedia.org/wiki/Ajax_(programming)}, último acesso em 29/03/2021}, o JavaScript começou a se comunicar com o servidor e com isso passou a realizar tarefas mais importantes nas aplicações web. Atualmente é raro encontrar aplicações que não utilizem JavaScript para funcionar.

Devido a popularização da linguagem, algumas dificuldades dos desenvolvedores começaram a ficar mais evidentes, como a necessidade de manipulação do DOM\footnote{\url{https://en.wikipedia.org/wiki/Document_Object_Model}, último acesso em 29/03/2021} (Document Object Model) e da interação do usuários através de eventos, por exemplo: o clique do mouse. As funções já existentes não possuíam uma solução eficiente para tais necessidades e começaram a surgir frameworks como Angular, Vue e React que fornecem um caminho para suprir essas dificuldades.

\subsubsection{Angular}

AngularJS\footnote{\url{https://en.wikipedia.org/wiki/AngularJS}}, depois apenas Angular, foi criado em 2010 por uma equipe de desenvolvedores da Google e foi o pioneiro na popularização de frameworks em JavaScript. Ele surge focado no desenvolvimento de SPAs (Single Page Applications), com uma arquitetura MVC (Model-View-Controller), separando a lógica do aplicativo, lógica de exibição e modelos.

O Angular permite que o desenvolvedor possa estender o HTML criando novas tags que podem ser reutilizada em outra parte do código quando necessário. Outra característica importante que ficou conhecido por \textit{two-way data-bindings}, vinculava o modelo com o objeto JavaScript permitindo a atualização automática quando acontecia mudança no HTML ou no próprio modelo.

\subsection{Vue}

Vue\footnote{\url{https://vuejs.org/}, último acesso em 29/03/2021} é bastante parecido com Angular,  foi criado por Evan You depois de trabalhar na Google utilizando Angular com o objetivo de ser um framework mais simples de se utilizar e na redução do número de arquivos usados. 

Na documentação oficial, Vue é definido como um framework progressivo para construção de interfaces de usuário. Em suas  últimas versões já apresenta uma abordagem baseada em componentes e um DOM virtual de maneira similar ao React.

O DOM virtual é uma representação de um objeto DOM. Um objeto DOM virtual possui as mesmas propriedades de um objeto DOM, no entanto manipulá-lo é muito mais rápido. 

\subsubsection{React}

Na documentação oficial, React\footnote{\url{https://reactjs.org/}, último acesso em 29/03/2021} é definido como uma biblioteca JavaScript declarativa, eficiente e flexível para criar interfaces com o usuário através de componentes. 

React foi criado pelo Facebook e o primeiro framework a usar um DOM virtual. Em relação a sua estrutura, React não utiliza o conceito de MVC, sua visualização é representada como uma árvore de entidades chamadas componentes que podem ser compostos de outros componentes.

\subsection{Desenvolvimento para aplicações móveis}

Depois de dominar o mundo da web, o JavaScript começou a ser utilizado como uma alternativa para lidar com o desenvolvimento multiplataforma em dispositivos móveis. Aplicativos híbridos, como ficou conhecido, é a combinação de tecnologias da web como HTML, o próprio JavaScript e CSS, Cascading Style Sheets, para construção de aplicativos móveis. 

\begin{figure}[ht]
    \centering
    \includegraphics[scale=.7]{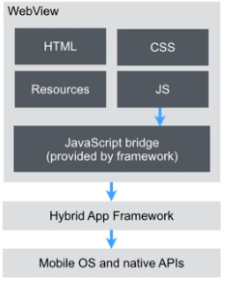}
    \caption{Abstração da arquitetura de uma aplicação híbrida}
    \label{fig:apphibrida}
\end{figure}

Os aplicativos híbridos são hospedados em uma aplicação nativa executando em cima de uma WebView, que é o nome dado ao browser encapsulado dentro da aplicação. Isso permite o acesso aos recursos do dispositivo como câmera, contatos, geolocalização etc. A Figura \ref{fig:apphibrida} apresenta uma abstração da arquitetura de um aplicativo híbrido.

Assim como ocorreu na web, começaram a surgir \textit{frameworks} JavaScript focados no desenvolvimento de aplicações híbridas como o PhoneGap/Apache Cordova entregando um conjunto de \textit{plugins} que auxiliam o acesso aos recursos do aparelho. O PhoneGap\footnote{\url{https://phonegap.com/}, último acesso em 29/03/2021} foi criado em 2009 pela empresa Nitobi, porém em 2011 a Adobe comprou a empresa e doou o código para o projeto Apache. Depois desse acontecimento, nasceu o Apache Cordova\footnote{\url{https://cordova.apache.org/}, último acesso em 29/03/2021} projeto \textit{open source} mantido por sua comunidade. 

A principal limitação de uma aplicação híbrida é quando se fala em performance. Atualmente a abordagem híbrida não consegue ser tão performática que uma aplicação nativa, justamente pelo fato do poder de renderização da \textit{WebView}. Quando falo em aplicações nativas significa que o desenvolvimento da aplicação foi realizado com a linguagem de programação padrão do SDK (\textit{Software Developer's Kit}) do dispositivo. No caso do Android temos o Java e, mais recentemente, Kotlin e para iOS temos o Objective-C e Swift. 

Todavia, a aplicação híbrida possui como maior ponto positivo a capacidade de, através de um único código fonte, entregar a aplicação em múltiplas plataformas, trazendo consigo uma série de benefícios como: velocidade de desenvolvimento e facilidade de manutenção do código. 

Diante desse cenário, temos o surgimento do React Native como uma solução que consegue melhorar o problema de performance da aplicação híbrida e mantém a prática de, através de um único código, entregar a aplicação nas principais plataformas móveis. 

\subsubsection{React Native}

O React Native \cite{ReactNative}, assim como o React, foi criado e é mantido pelo Facebook. React Native utiliza o mesmo design do React, permitindo compor ricas interfaces de usuário a partir de componentes declarativos.

\begin{figure}[ht]
    \centering
    \includegraphics[scale=0.8]{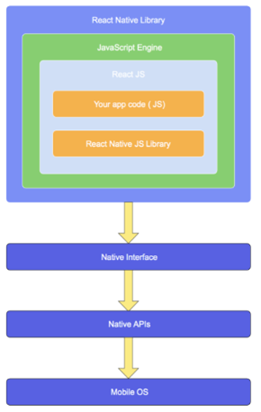}
    \caption{Abstração da arquitetura de uma aplicação em React Native}
    \label{fig:appreactnative}
\end{figure}

Com React Native, você não cria um aplicativo híbrido utilizando as ferramentas oriundas da web, no entanto é possível criar um aplicativo móvel real que é indistinguível de um aplicativo nativo, pois ele usa os mesmos blocos de construção fundamentais da interface do usuário dos aplicativos iOS e Android comuns. Diferente de ferramentas como PhoneGap/Apache Cordova que executa o JavaScript e permitir sua interação com recursos nativos, o React Native compila o JavaScript ao código nativo, explicando assim como ele consegue melhorar a performance da aplicação. A Figura \ref{fig:appreactnative} apresenta uma abstração da arquitetura do React Native. React Native vem crescendo junto com sua comunidade e sendo adotado por grandes empresas como: Instagram, Uber e o próprio Facebook. 

Os principais problemas associados ao React Native são de gerenciamento de configuração, dependências e testes. Por ser apenas um código, na teoria se um teste passou no ambiente iOS, podemos assumir que o teste também passará no ambiente Android. Na maioria dos casos o cenário descrito acima é verídico, que é testemunho do poder do React Native, porém existem exceções suficientes para causar problemas no ciclo de controle e garantia de qualidade do aplicativo, como por exemplo: bibliotecas que funcionam bem para iOS, porém não no Android. 

Diante desse cenário, uma abordagem para solucionar tais problemas é o desenvolvimento e utilização de uma pipeline DevOps. DevOps se propõe a resolver esse problema combinando a mudança cultural com a automação de entrega de software com objetivo de garantir lançamentos de software frequentes, previsíveis e sem erros. 

\subsection{DevOps}

DevOps é um conjunto de práticas destinadas a reduzir o tempo entre fazer uma alteração em um sistema e a mudança ser colocada em produção normal, garantindo alta qualidade \cite{BassWeberZhu15}.
	
\begin{figure}[ht]
    \centering
    \includegraphics[width=\linewidth]{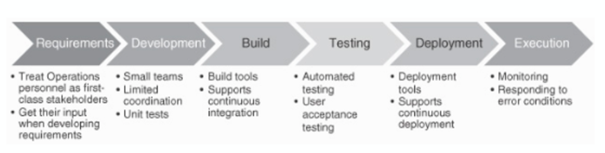}
    \caption{Abstração de uma pipeline DevOps}
    \label{fig:pipelinedevops}
\end{figure}

DevOps surgiu na engenharia de software com ênfase em colaboração, automação e utilização de ferramentas como ponte de comunicação entre as atividades de desenvolvimento e operações, em resposta a crescente utilização das metodologias ágeis e ao aumento da demanda de produção de aplicativos de software, que necessitavam de entregas rápidas, contínuas e com mais qualidade. A Figura \ref{fig:pipelinedevops} apresenta uma abstração de uma pipeline DevOps.

\begin{figure}[ht]
    \centering
    \includegraphics[width=\linewidth]{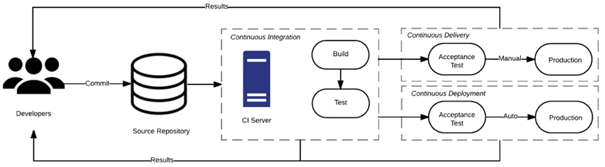}
    \caption{Relacionamento entre integração contínua, entrega contínua e implantação contínua \cite{Shahin2017a}}
    \label{fig:continuous}
\end{figure}

DevOps se inclui área emergente de pesquisa e prática chamada de Engenharia de Software Contínua. Essa área refere-se ao desenvolvimento, implantação e obtenção de \textit{feedback} rápido do software e do cliente em um ciclo muito rápido \cite{Bosch14, Fitzgerald2017}. Engenharia de Software Contínua envolve três fases: Estratégia de Negócio e Planejamento, Desenvolvimento e Operações. Para este estudo iremos focar nas atividades relacionadas a fase de Desenvolvimento: Integração contínua, Entrega contínua e Implantação contínua. A Figura \ref{fig:continuous} apresenta o relacionamento entre essas atividades.

\subsubsection{Integração contínua}

Integração contínua é um conjunto de práticas de desenvolvimento de software, no qual os desenvolvedores publicam regularmente as alterações em seus códigos com o propósito de reduzir o tempo para disponibilizar novas atualizações no software \cite{Fowler2006Blog}.

Martin Fowler define integração contínua como: ``\emph{uma prática de desenvolvimento de software em que os membros de uma equipe integram seu trabalho com frequência, geralmente cada pessoa realiza pelo menos uma integração diariamente - levando a múltiplas integrações por dia. Cada integração é verificada por uma compilação automatizada (incluindo teste) para detectar erros de integração o mais rápido possível}'' \cite{Fowler2006Blog}.

A utilização dessa prática permite que as empresas de software tenham um ciclo de lançamento mais curto e frequente, melhorem a qualidade do software e aumentem a produtividade de suas equipes. Esta prática inclui a construção de um servidor com uma ferramenta de integração contínua como o Jenkins e implementação de testes automatizado. 

\subsubsection{Entrega contínua}

Segundo \cite{Humble2010}, entrega contínua é a capacidade de obter mudanças de todos os tipos - incluindo novos recursos, alterações de configuração, correções de erros e experimentos - na produção ou nas mãos dos usuários, com segurança e rapidez de maneira sustentável.

Entrega contínua nasce com o objetivo de substituir o processo de entrega de software tradicional e tal prática é possível garantindo um bom processo de integração contínua, deixando o software em um estado entregável após cada mudança efetivada \cite{Weber2016} após passar com sucesso em testes automatizados e verificações de qualidade. 

De acordo com \cite{Chen2015a}, esta prática oferece bastante benefícios como como redução do risco de implantação, redução de custos, aumento da satisfação do usuário, produtividade e eficiência. Conforme apresentado na Figura \ref{fig:continuous} esta prática tem como requisito a utilização de integração contínua. 

\subsubsection{Implantação contínua}

Implantação Contínua é uma prática de desenvolvimento de software na qual cada alteração de código passa por toda a pipeline e é colocada em produção, automaticamente. 

Não existe um consenso no meio acadêmico sobre a definição e diferença entre Entrega contínua e Implantação contínua \cite{Fitzgerald2017}, no entanto é importante ressaltar que a prática de Implantação contínua implica obrigatoriamente na prática de Entrega contínua, mas o inverso não é verdadeiro \cite{Humble2010Blog}. 

\begin{figure}[ht]
    \centering
    \includegraphics[width=\linewidth]{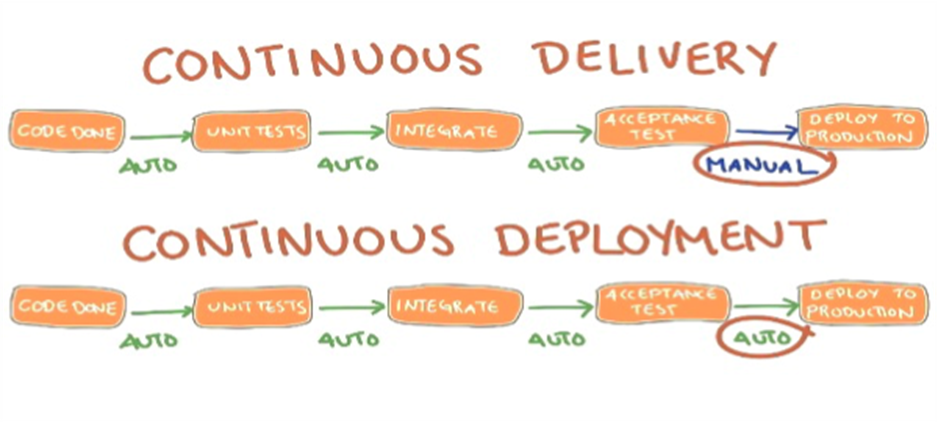}
    \caption{Diferença entre Entrega contínua e Implantação contínua \cite{Humble2010Blog}}
    \label{fig:delivery-deploy}
\end{figure}

A principal diferença entre Entrega contínua e Implantação contínua está no momento da entrega do software. Na Entrega contínua, seu software estará sempre em um estado entregável após as mudanças efetivadas, porém o momento de distribuição para o cliente final é um passo manual, ou seja, uma decisão de negócio. No entanto, conforme apresentado na Figura \ref{fig:delivery-deploy}, na Implantação contínua o processo é totalmente automatizado.

\subsection{Discussão}

Nesta Seção foi apresentado todo um contexto teórico e tecnológico para termos uma noção sobre os benefícios e desafios associados à implantação de uma pipeline DevOps. 

A próxima Seção apresenta como foi realizado o desenvolvimento do aplicativo bem como a implementação de seus testes.

\section{Implementação da proposta}

Nesta seção é apresentada a escolha pela tecnologia adotada para o desenvolvimento do aplicativo, bem como sua implementação e os \textit{scripts} de testes elaborados para o propósito deste estudo.

\subsection{Tecnologia adotada}

Conforme apresentado na Seção \ref{subsec:frameworks}, na Seção de conceitos básicos, React Native é uma biblioteca para construção de aplicativos móveis. Pelo fato desta tecnologia ter JavaScript em seu núcleo, é muito mais fácil encontrar desenvolvedores em comparação a outras linguagens como Swift ou Kotlin.

\begin{figure}[ht]
    \centering
    \includegraphics[width=\linewidth]{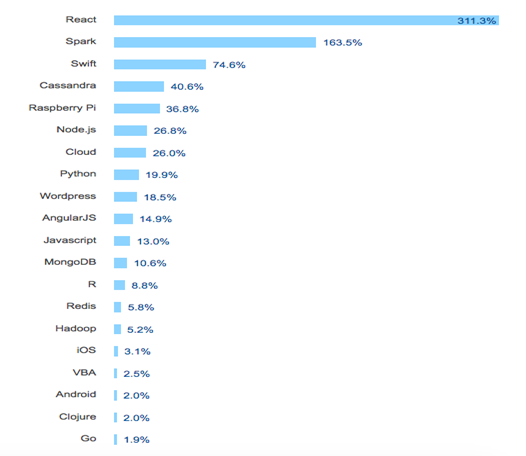}
    \caption{Alteração na parcela de votos do Stack Overflow entre janeiro de 2015 e janeiro de 2016 \cite{StackOverflowDeveloperSurvey2016}}
    \label{fig:stackoverflow}
\end{figure}

A Stack Overflow , maior comunidade online de desenvolvedores do mundo, publica todo ano um \textit{survey} apresentando um consolidado das respostas de seus usuários sobre vários temas, desde tecnologias favoritas até preferências de trabalho. No seu \textit{survey} \cite{StackOverflowDeveloperSurvey2018} publicado neste ano de 2018, JavaScript atingiu a marca de linguagem de programação mais popular do mundo pelo sexto ano consecutivo. O \textit{survey} publicado em 2016 \cite{StackOverflowDeveloperSurvey2016} apontou React como a principal tendência tecnológica no contexto de desenvolvimento como ilustrado na Figura \ref{fig:stackoverflow}, a confirmação desta tendência está presente no \textit{survey} deste ano no qual React aparece na terceira colocação na categoria de Frameworks, Libraries and Tools mais utilizada atualmente. 

\subsection{Implementação dos testes}
\label{sec:testes}

Brian Marick, em uma série de \textit{posts}\footnote{\url{https://codegardener.com/tag/Brian\%20Marick}, último acesso em 29/03/2021}, fez uma classificação de testes de software em uma forma chamada \textit{The Marick Test Matrix}. De acordo com Marick, podemos dividir a área de testes em quatro tipos, são elas: testes de aceitação, unitário, exploratório e não funcional.

Neste estudo, para atender o objetivo, focamos na implementação apenas dos testes de aceitação que representa requisitos funcionais vistos da perspectiva de negócio. Geralmente, esse tipo de teste, é escrito na forma de histórias de usuários exemplificando como o software deve funcionar. O aplicativo  desenvolvido para este estudo, consiste em uma listagem de professores e para o mesmo foram definidos duas histórias de usuários:

\begin{itemize}
    \item Como usuário, eu quero visualizar a lista de professores
    \item Como usuário, eu quero visualizar os detalhes de um professor
\end{itemize}

\begin{figure}[ht]
    \centering
    \includegraphics[width=\linewidth]{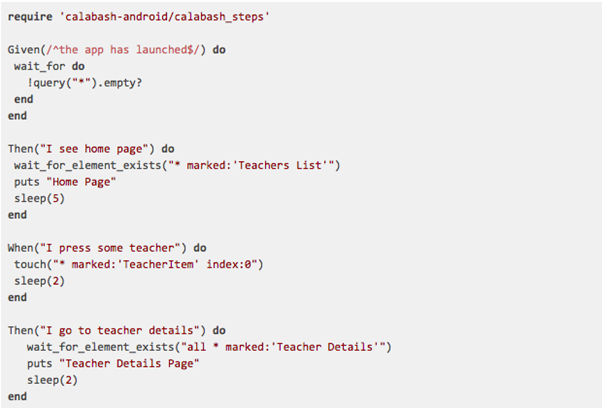}
    \caption{Código do teste de aceitação}
    \label{fig:codigotesteaceitacao}
\end{figure}

Os testes presentes na Figura \ref{fig:codigotesteaceitacao} foram implementados utilizando a ferramenta \textit{open source} Calabash\footnote{\url{https://calaba.sh/}, último acesso em 29/03/2021} em sua versão 0.9.7. Calabash permite a escrita e execução de testes de aceitação automatizados de aplicativos móveis. Ele consiste em uma biblioteca que permite o código de teste interagir com o aplicativo simulando ações do usuário final.

As palavras reservadas \texttt{Given}, \texttt{When} e \texttt{Then} são expressões do Cucumber\footnote{\url{https://cucumber.io/}, último acesso em 29/03/2021}, ferramenta que permite a utilização de linguagem natural para expressar o comportamento do usuário ao utilizar o aplicativo. No trecho de código abaixo estamos descrevendo a ação do usuário de abrir a aplicação.

\begin{verbatim}
Given(/^the app has launched$/) do
    wait_for do
        !query("*").empty?
    end
end
\end{verbatim}

No entanto, as expressões \texttt{wait\_for}, \texttt{query}, \texttt{sleep}, \texttt{touch} e  \texttt{wait\_for\_element\_exists} são palavras reservadas do Calabash responsáveis por simular as ações do usuário interagindo com a aplicação. No trecho de código abaixo estamos simulando a ação de clique do usuário em um item da listagem de professores. 

\begin{verbatim}
When("I press some teacher") do
  touch("* marked:'TeacherItem' index:0")
  sleep(2)
end
\end{verbatim}

\subsection{Implementação do aplicativo}

\begin{figure}[ht]
    \centering
    \includegraphics[width=\linewidth]{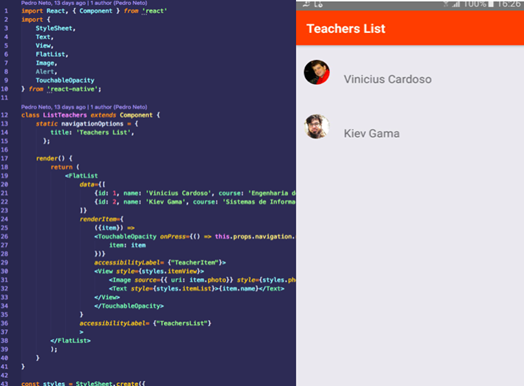}
    \caption{Código e a tela de listagem de professores}
    \label{fig:screenlistteachers}
\end{figure}

\begin{figure}[ht]
    \centering
    \includegraphics[width=\linewidth]{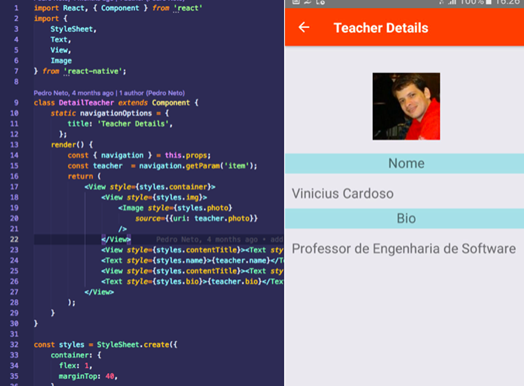}
    \caption{Código e tela do detalhe de professor}
    \label{fig:screenteacherdetails}
\end{figure}

Após a definição das histórias de usuários e a elaboração dos casos de teste foi iniciado a fase de implementação do aplicativo em React Native na versão 0.55.2. A Figura \ref{fig:screenlistteachers} apresenta o código e a tela de listagem dos professores e a Figura \ref{fig:screenteacherdetails} apresenta o código e a tela de detalhe de professor. Todo código desenvolvido está disponível em um repositório público\footnote{\url{https://github.com/pedrojsn96/intro-react/tree/master/TeachersProject}, último acesso em 29/03/2021} no Github. 

O trecho de código abaixo ilustra a classe responsável pela tela de listagem dos professores. 

\begin{figure}[ht]
    \centering
    \includegraphics[width=\linewidth]{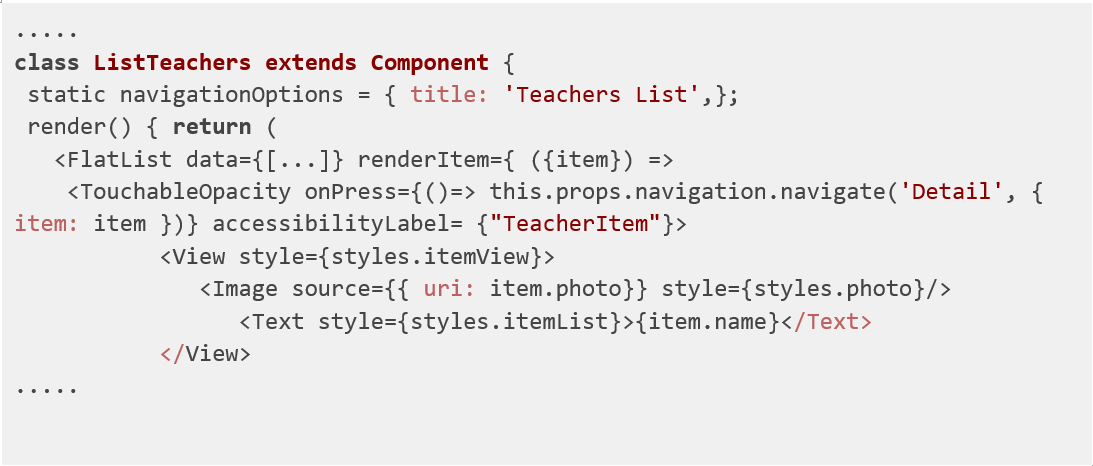}
    \label{fig:sourcecode1}
\end{figure}

\subsection{Discussão}

Nesta Seção foi apresentada a justificativa da tecnologia adotada, a implementação dos testes e o desenvolvimento do aplicativo em React Native que será utilizado na \textit{pipeline} DevOps.

A implementação do aplicativo foi importante para sentir as dificuldades de um desenvolvedor React Native com o gerenciamento de versões de dependências, compreender o motivo pelo qual React é o framework JavaScript mais utilizado atualmente e  entender as dificuldades e limitações das ferramentas disponíveis para testar esse tipo de aplicação. 

A próxima Seção apresenta como foi o passo a passo do processo de construção da pipeline de integração e entrega contínua.

\section{A Pipeline de Integração e Entrega Contínua para aplicações móveis desenvolvidas em React Native}

Rafał Leszko \cite{Leszko2017CdwD} apresenta o conceito de pipeline como ``\emph{uma sequência de operações automatizadas que geralmente representa uma parte da entrega de software e o processo de garantia de qualidade}''. 

O sucesso da implantação de uma pipeline de integração e entrega contínua não depende exclusivamente das ferramentas que vão ser utilizadas, mas principalmente da estrutura da organização, equipe de desenvolvimento e as suas práticas. 

\begin{figure}[ht]
    \centering
    \includegraphics[width=\linewidth]{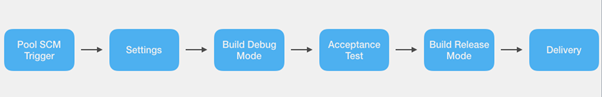}
    \caption{Pipeline desenvolvida para este trabalho}
    \label{fig:mypipeline}
\end{figure}

A pipeline pode variar de acordo com o aplicativo em questão, no sentido de possuir mais ou menos passos para atender a necessidade do mesmo. A Figura \ref{fig:mypipeline} apresenta a pipeline para o aplicativo desenvolvido neste estudo.

\subsection{Configuração de ambiente}
\label{sec:configuracao}

No processo de implementação de uma pipeline de integração e entrega contínua é necessário fazer uso de ferramentas para dar suporte ao controle de versionamento de código, automação de \textit{build}, automação de testes etc. Neste estudo utilizamos o Github\footnote{\url{https://github.com/}, último acesso em 29/03/2021} como ferramenta para versionamento de código, Jenkins na versão 2.138.2 como servidor de integração contínua e o Gradle\footnote{\url{https://gradle.org/}, último acesso em 29/03/2021} para realização do \textit{build}.

\subsubsection{Jenkins}

Na configuração do servidor Jenkins foram instalados os plugins  Blue Ocean na versão 1.9.0, Slack Notification Plugin na versão 2.3, TestFairy Plugin na versão 4.16 e  Github Integration Plugin na versão 0.2.4 para auxiliar na integração e comunicação entre o Jenkins e as demais aplicações utilizadas no desenvolvimento da pipeline. 

\section{Estrutura da pipeline}

Ao lidar com pipeline no Jenkins é preciso ter em mente dois elementos: etapa e passo. 

\begin{itemize}
    \item \textbf{Passo}: É a operação mais básica, uma tarefa única onde se diz o que o Jenkins vai fazer. Por exemplo: executar um \textit{script}.
    \item \textbf{Etapa}: É um conjunto de passos, dentro do mesmo contexto, que serão executados de maneira sequencial ou paralela. Por exemplo: sequência de \textit{scripts} de configuração. 
\end{itemize}

\begin{figure}[ht]
    \centering
    \includegraphics[width=\linewidth]{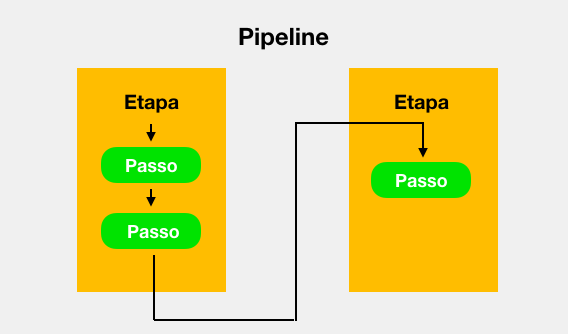}
    \caption{Estrutura de pipeline}
    \label{fig:ypipelinestructure}
\end{figure}

A Figura \ref{fig:ypipelinestructure} apresenta uma visão geral da relação entre etapa e passo.

\subsection{Jenkinsfile}

Para construção de pipeline no Jenkins é necessário escrever um arquivo de texto chamado de \texttt{Jenkinsfile}. Neste arquivo contém a declaração de todas as etapas e passos presentes na sua pipeline. 

O \texttt{Jenkinsfile} fica na raiz do seu projeto sendo considerado como parte do desenvolvimento. Esta prática deu origem ao termo \emph{"pipeline-as-code"} que trata a pipeline de integração e entrega contínua como parte do aplicativo que precisa ser versionado e revisado. 

\subsection{Poll SCM Trigger}

A primeira etapa é responsável por dar início a execução da \textit{pipeline}. Ela é acionada automaticamente, através de \texttt{cron job}\footnote{\url{https://en.wikipedia.org/wiki/Cron}, último acesso em 29/03/2021} que executa a cada doze horas de segunda a sexta verificando a existência de alterações no repositório. Caso haja de novos \textit{commits} no repositório, o processo de pipeline é iniciado. 

O bloco de código a seguir apresenta a parte do \texttt{Jenkinsfile} responsável por esta etapa.

\begin{figure}[ht]
    \centering
    \includegraphics[width=\linewidth]{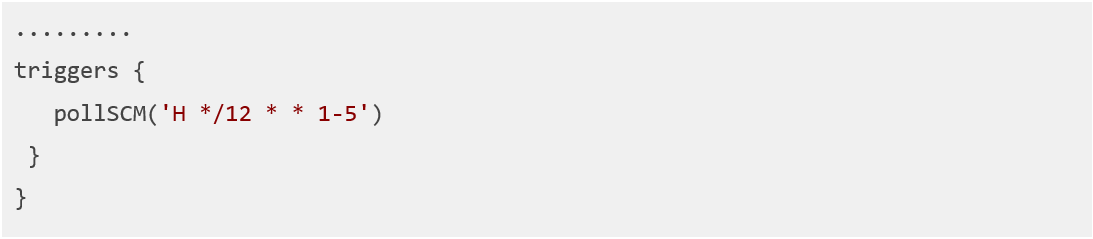}
    \label{fig:sourcecode2}
\end{figure}

\subsection{Settings}

A etapa de \textit{Settings} contém duas sub-etapas chamadas de \texttt{BUNDLE INSTALL} e \texttt{NPM INSTALL}. Esta etapa é responsável pelo gerenciamento das dependências do projeto. Durante a execução da pipeline as duas sub-etapas são executadas em paralelo.

A sub-etapa de \texttt{BUNDLE INSTALL} garante que todas as dependências necessárias para testar a aplicação estejam devidamente instaladas e versionadas. Enquanto a sub-etapa de \texttt{NPM INSTALL} realiza a instalação de todas as dependência, devidamente versionadas, que a aplicação necessita para funcionar corretamente. 

O bloco de código a seguir apresenta a parte do \texttt{Jenkinsfile} responsável por esta etapa.

\begin{figure}[ht]
    \centering
    \includegraphics[width=\linewidth]{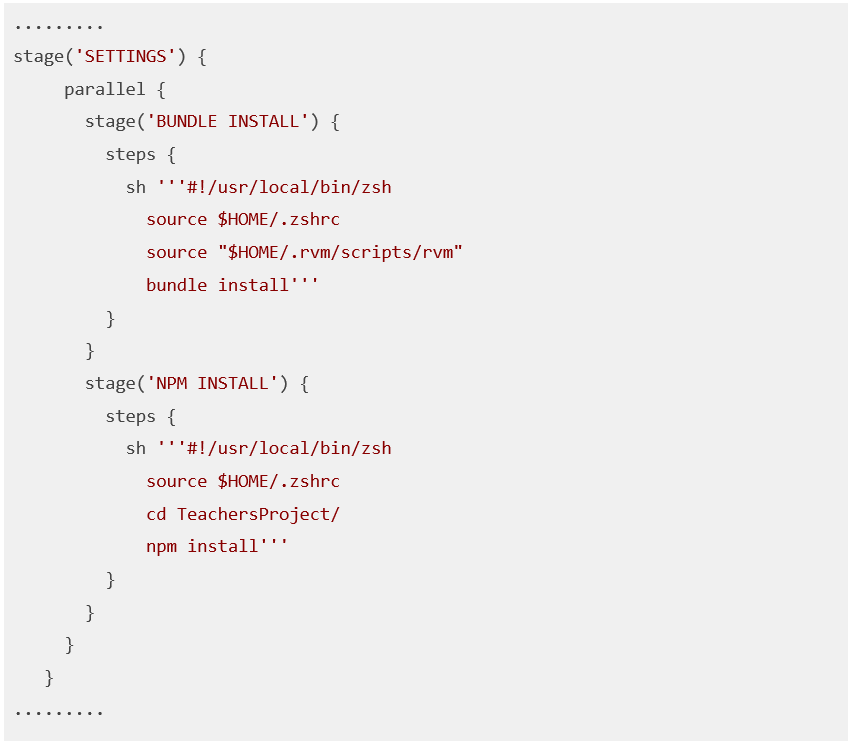}
    \label{fig:sourcecode3}
\end{figure}

\subsection{Build Debug Mode}

Esta etapa é responsável por criar o \textit{build} no modo de \textit{debug}, neste modo, por exemplo, é possível visualizar todos os erros e alertas gerados pela aplicação em tempo de execução. 

\begin{figure}[ht]
    \centering
    \includegraphics[width=\linewidth]{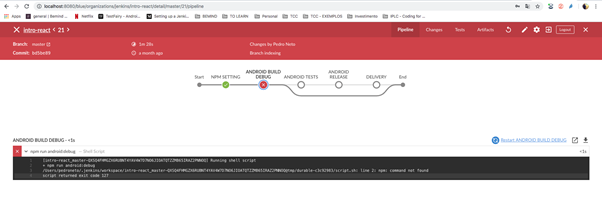}
    \caption{Erro durante a execução da pipeline}
    \label{fig:pipelineerror}
\end{figure}

Ainda nesta etapa, um arquivo \texttt{apk} (\textit{Android Application Package}), é gerado sendo possível executar os cenários de testes na próxima etapa da pipeline. Para esse passo executamos um \textit{script} que aciona o Gradle, ferramenta responsável pelo \textit{build} da aplicação conforme apresentado na Seção \ref{sec:configuracao} Se alguma coisa der errado, a execução da \textit{pipeline} é interrompida apresentando visualmente em qual passo o erro aconteceu, como ilustrado na Figura \ref{fig:pipelineerror}, caso contrário, seguirá para a próxima etapa. 

O bloco de código a seguir apresenta a parte do \texttt{Jenkinsfile} responsável por esta etapa.

\begin{figure}[ht]
    \centering
    \includegraphics[width=\linewidth]{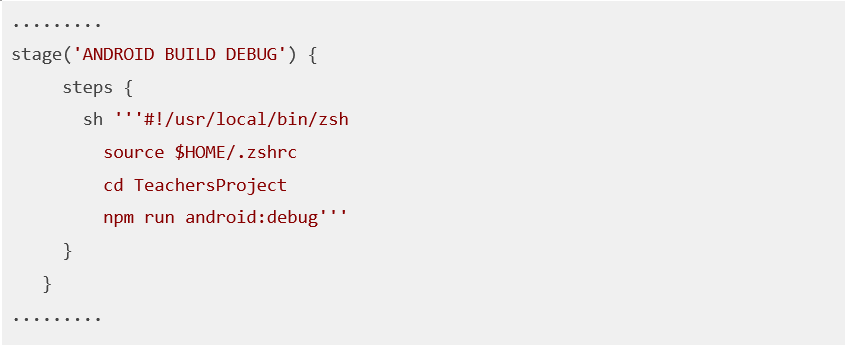}
    \label{fig:sourcecode4}
\end{figure}

\subsection{Teste de Aceitação}

A etapa de Teste de Aceitação é a responsável por executar os testes do aplicativo, apresentados na Seção \ref{sec:testes}. Esta etapa é importante, pois é ela que vai garantir a qualidade do software e decidirá se o aplicativo está pronto ou não para ser entregue para os usuários. Se acontecer erro em algum dos testes, a execução da pipeline é interrompida, caso contrário, o aplicativo seguirá para próxima etapa.

Para esta etapa obter sucesso, existe a necessidade de possuir um dispositivo Android, configurado em modo de desenvolvimento, conectado via \textit{usb} na máquina ou a utilização de um simulador Android. O aplicativo será instalado automaticamente no dispositivo e os testes serão executados no mesmo.

O bloco de código a seguir apresenta a parte do \texttt{Jenkinsfile} responsável por esta etapa.

\begin{figure}[ht]
    \centering
    \includegraphics[width=\linewidth]{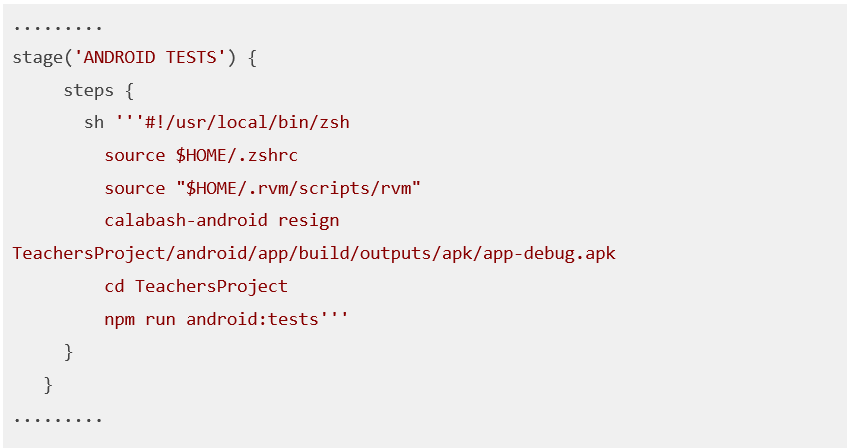}
    \label{fig:sourcecode5}
\end{figure}

\subsection{Build Release Mode}

Esta etapa tem a função de preparar o aplicativo para ser enviado para Google Play Store\footnote{\url{https://play.google.com/store}, último acesso 29/03/2021}. O Android exige que todos os aplicativos sejam assinados digitalmente com um certificado antes de poderem ser distribuídos e instalados.

\begin{figure}[ht]
    \centering
    \includegraphics[width=\linewidth]{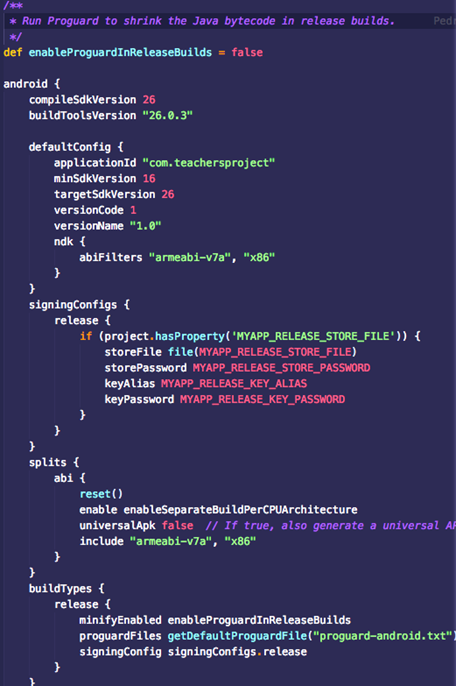}
    \caption{Assinatura do aplicativo para distribuição na loja Google Play}
    \label{fig:googleplay}
\end{figure}

Para assinar digitalmente é necessário seguir a documentação oficial da plataforma\footnote{\url{https://facebook.github.io/react-native/docs/signed-apk-android}, último acesso em 29/03/2021} e adicionar o certificado gerado ao seu código, como pode ser visto na Figura \ref{fig:googleplay}. Dado que o certificado já foi devidamente inserido ao código, agora é possível gerar um \textit{build} no modo de \textit{release}.

Para gerar o \textit{build} no modo de \textit{release}, executamos um \textit{script} que aciona novamente o Gradle. Se ocorrer erro durante a execução deste passo, a pipeline é interrompida, caso contrário, seguirá para a próxima etapa.

O bloco de código a seguir apresenta a parte do \texttt{Jenkinsfile} responsável por esta etapa.

\begin{figure}[ht]
    \centering
    \includegraphics[width=\linewidth]{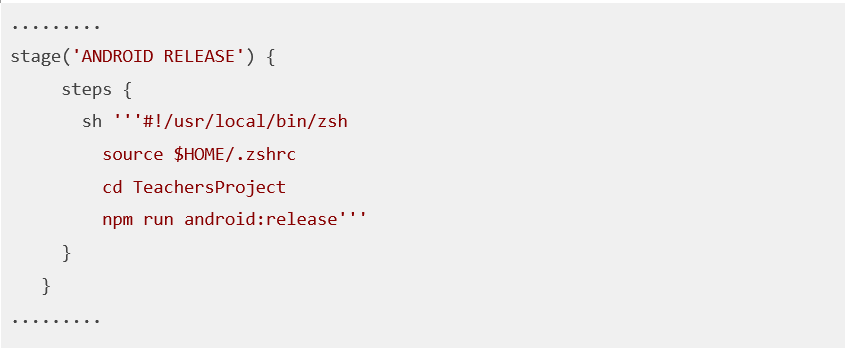}
    \label{fig:sourcecode6}
\end{figure}

\subsection{Entrega}

A última etapa é responsável basicamente por uma tarefa: \textbf{notificação}. No nosso caso, esta etapa notifica dois públicos distintos: o time de desenvolvimento e grupo de usuários que receberam o aplicativo para testar.  

\begin{figure}[ht]
    \centering
    \includegraphics[width=\linewidth]{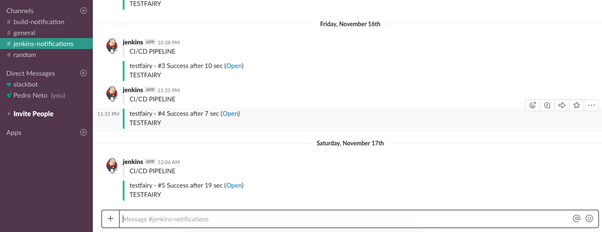}
    \caption{Notificação para o time de desenvolvimento}
    \label{fig:slack}
\end{figure}

O time de desenvolvimento recebe o aviso através de um \textit{bot} do Jenkins em um canal no Slack indicando que a execução da pipeline foi finalizada, conforme a Figura \ref{fig:slack}. 

\begin{figure}[ht]
    \centering
    \includegraphics[width=\linewidth]{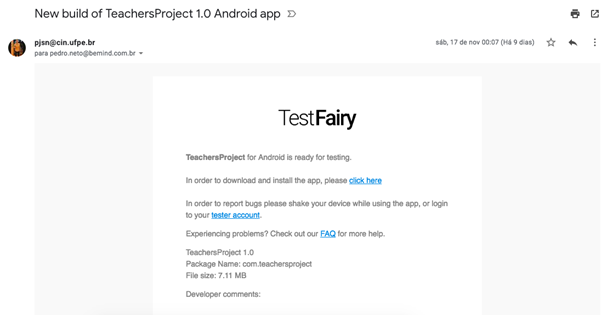}
    \caption{Notificação para o grupo de usuários}
    \label{fig:usersnotification}
\end{figure}

Por sua vez, o grupo de usuários recebe o aviso através de um e-mail contendo os devidos links para realização do download do aplicativo, conforme a Figura \ref{fig:usersnotification}.

O bloco de código a seguir apresenta a parte do \texttt{Jenkinsfile} responsável por esta etapa.

\begin{figure}[ht]
    \centering
    \includegraphics[width=\linewidth]{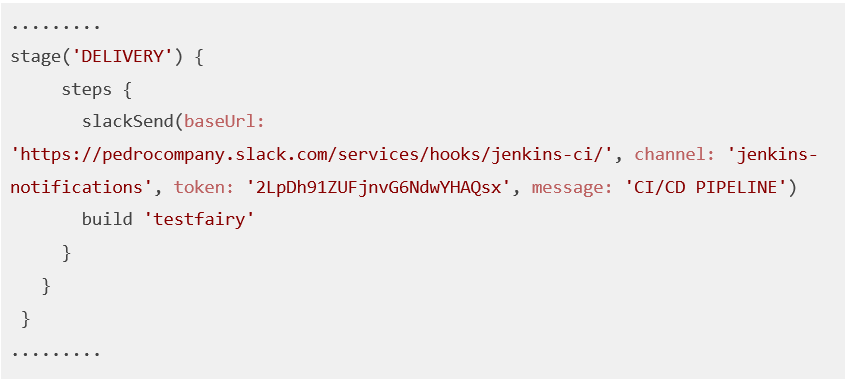}
    \label{fig:sourcecode7}
\end{figure}

\subsection{Discussão}

Nesta Seção foi apresentado o conceito, a estrutura e a pipeline desenvolvida para o estudo.  Bem como as ferramentas utilizadas para dar suporte a integração contínua, ao versionamento de código, ao \textit{build} e a entrega do aplicativo. É apresentado o detalhe de cada etapa da pipeline explicando suas responsabilidades e funcionamento. 

\begin{figure}[ht]
    \centering
    \includegraphics[width=\linewidth]{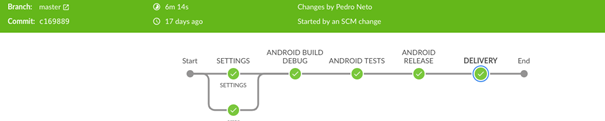}
    \caption{Todas as etapas da pipeline após a execução}
    \label{fig:successfulpipeline}
\end{figure}

A etapa que apresentou maior dificuldade de ser implementada foi a de Testes de Aceitação, pelo fato da mesma ser responsável por garantir a qualidade do software que será entregue para os usuários e talvez seja um dos principais motivos para quebrar o processo de automação do pipeline por completo. A Figura \ref{fig:successfulpipeline} apresenta todas as etapas da pipeline após a execução.

A próxima Seção apresenta as conclusões, limitações e os trabalhos futuros necessários para cobrir toda a fase de Desenvolvimento da Engenharia de Software Contínua. 

\section{Considerações Finais}

A construção de uma pipeline fornece um mecanismo para organização do processo de desenvolvimento de software como um todo, automatizando \textit{build} e a maneira como o software é entregue para o cliente. Com a popularização no uso de integração e entrega contínua, a construção de pipelines estará presente no cotidiano de toda empresa de desenvolvimento nos próximos anos.

Com o desenvolvimento deste trabalho, foi possível provar com meios práticos e utilização de ferramentas testadas no mercado e amplamente conhecidas no ambiente organizacional que é tangível a utilização de integração e entrega contínua, altamente difundida no desenvolvimento web, agregando assim todos os benefícios já conhecidos atrelados a utilização desta prática, para o ambiente de desenvolvimento de aplicativos móveis.

\subsection{Dificuldades e lições aprendidas}

Neste estudo a maior dificuldade encontrada foi que o React Native possui um conjunto limitado de ferramentas para realizar testes. Calabash, a ferramenta escolhida, possui uma pequena comunidade de usuários o que dificulta para encontrar soluções de possíveis problemas e apresenta dificuldades para inspecionar e identificar os elementos na tela, para contornar esse problema é necessário adicionar a propriedade do React Native accessibilityLabel, que é uma propriedade que torna os aplicativos acessíveis para pessoas com deficiências indicando que o elemento está visível para ser manipulado. Esta propriedade auxilia na busca por elementos pelas ferramentas de testes, além de ser uma boa prática de desenvolvimento.  

\subsection{Trabalhos futuros e limitações}

Neste estudo não foi realizado a implementação de testes unitários e não foi aplicado uma ferramenta de Cobertura de Testes, na qual se exige uma porcentagem mínima de código coberto por teste para  um \texttt{commit} ser aceito e dar início a execução da pipeline. Ficamos limitado ao ambiente Android, não foi possível incluir o processo de \textit{build} para iOS, bem como não foi implementado o passo da implantação contínua responsável, no contexto de aplicação móvel, por publicar o aplicativo nas lojas.

Como principais atividades futuras esta implementação dos testes unitários utilizando Jest\footnote{\url{https://jestjs.io/}, último acesso em 29/03/2021}, ferramenta utilizada para testar aplicações desenvolvidas em React, integrar ao pipeline a ferramenta Fastlane\footnote{\url{https://fastlane.tools/}, último acesso em 29/03/2021} para auxiliar na etapa de \textit{build} da aplicação para o ambiente iOS e no processo de publicação tanto na Google Play Store quanto na Apple Store. Com essas atividades será possível cobrir toda a fase de Desenvolvimento da Engenharia de Software Contínua.

\bibliographystyle{ACM-Reference-Format}
\bibliography{acmart}


\begin{thebibliography}{19}


\ifx \showCODEN    \undefined \def \showCODEN     #1{\unskip}     \fi
\ifx \showDOI      \undefined \def \showDOI       #1{#1}\fi
\ifx \showISBNx    \undefined \def \showISBNx     #1{\unskip}     \fi
\ifx \showISBNxiii \undefined \def \showISBNxiii  #1{\unskip}     \fi
\ifx \showISSN     \undefined \def \showISSN      #1{\unskip}     \fi
\ifx \showLCCN     \undefined \def \showLCCN      #1{\unskip}     \fi
\ifx \shownote     \undefined \def \shownote      #1{#1}          \fi
\ifx \showarticletitle \undefined \def \showarticletitle #1{#1}   \fi
\ifx \showURL      \undefined \def \showURL       {\relax}        \fi
\providecommand\bibfield[2]{#2}
\providecommand\bibinfo[2]{#2}
\providecommand\natexlab[1]{#1}
\providecommand\showeprint[2][]{arXiv:#2}

\bibitem[\protect\citeauthoryear{Bass, Weber, and Zhu}{Bass
  et~al\mbox{.}}{2015}]%
        {BassWeberZhu15}
\bibfield{author}{\bibinfo{person}{Len Bass}, \bibinfo{person}{Ingo Weber},
  {and} \bibinfo{person}{Liming Zhu}.} \bibinfo{year}{2015}\natexlab{}.
\newblock \bibinfo{booktitle}{\emph{DevOps: A Software Architect's
  Perspective}}.
\newblock \bibinfo{publisher}{Addison-Wesley}, \bibinfo{address}{New York}.
\newblock
\showISBNx{978-0-13-404984-7}
\urldef\tempurl%
\url{http://my.safaribooksonline.com/9780134049847}
\showURL{%
\tempurl}


\bibitem[\protect\citeauthoryear{Bosch}{Bosch}{2014}]%
        {Bosch14}
\bibfield{author}{\bibinfo{person}{Jan Bosch}.}
  \bibinfo{year}{2014}\natexlab{}.
\newblock \bibinfo{booktitle}{\emph{Continuous Software Engineering: An
  Introduction}}.
\newblock \bibinfo{publisher}{Springer International Publishing},
  \bibinfo{address}{Cham}, \bibinfo{pages}{3--13}.
\newblock
\showISBNx{978-3-319-11283-1}
\urldef\tempurl%
\url{https://doi.org/10.1007/978-3-319-11283-1_1}
\showDOI{\tempurl}


\bibitem[\protect\citeauthoryear{Chen}{Chen}{2015}]%
        {Chen2015a}
\bibfield{author}{\bibinfo{person}{Lianping Chen}.}
  \bibinfo{year}{2015}\natexlab{}.
\newblock \showarticletitle{{Continuous Delivery: Huge Benefits, but Challenges
  Too}}.
\newblock \bibinfo{journal}{\emph{IEEE Software}} \bibinfo{volume}{32},
  \bibinfo{number}{2} (\bibinfo{date}{mar} \bibinfo{year}{2015}),
  \bibinfo{pages}{50--54}.
\newblock
\showISBNx{9781467372848}
\showISSN{0740-7459}
\urldef\tempurl%
\url{https://doi.org/10.1109/MS.2015.27}
\showDOI{\tempurl}
\showeprint[arxiv]{arXiv:1011.1669v3}


\bibitem[\protect\citeauthoryear{Chen}{Chen}{2017}]%
        {Chen2017}
\bibfield{author}{\bibinfo{person}{Lianping Chen}.}
  \bibinfo{year}{2017}\natexlab{}.
\newblock \showarticletitle{{Continuous Delivery: Overcoming adoption
  challenges}}.
\newblock \bibinfo{journal}{\emph{Journal of Systems and Software}}
  \bibinfo{volume}{128} (\bibinfo{date}{jun} \bibinfo{year}{2017}),
  \bibinfo{pages}{72--86}.
\newblock
\showISSN{01641212}
\urldef\tempurl%
\url{https://doi.org/10.1016/j.jss.2017.02.013}
\showDOI{\tempurl}


\bibitem[\protect\citeauthoryear{Ebert, Gallardo, Hernantes, and Serrano}{Ebert
  et~al\mbox{.}}{2016}]%
        {Ebert2016}
\bibfield{author}{\bibinfo{person}{Christof Ebert}, \bibinfo{person}{Gorka
  Gallardo}, \bibinfo{person}{Josune Hernantes}, {and} \bibinfo{person}{Nicolas
  Serrano}.} \bibinfo{year}{2016}\natexlab{}.
\newblock \showarticletitle{{DevOps}}.
\newblock \bibinfo{journal}{\emph{IEEE Software}} \bibinfo{volume}{33},
  \bibinfo{number}{3} (\bibinfo{date}{may} \bibinfo{year}{2016}),
  \bibinfo{pages}{94--100}.
\newblock
\showISSN{0740-7459}
\urldef\tempurl%
\url{https://doi.org/10.1109/MS.2016.68}
\showDOI{\tempurl}


\bibitem[\protect\citeauthoryear{Facebook}{Facebook}{2020}]%
        {ReactNative}
\bibfield{author}{\bibinfo{person}{Facebook}.} \bibinfo{year}{2020}\natexlab{}.
\newblock \bibinfo{title}{React Native}.
\newblock
  \bibinfo{howpublished}{\url{https://github.com/facebook/react-native}}.
\newblock


\bibitem[\protect\citeauthoryear{Fitzgerald and Stol}{Fitzgerald and
  Stol}{2017}]%
        {Fitzgerald2017}
\bibfield{author}{\bibinfo{person}{Brian Fitzgerald} {and}
  \bibinfo{person}{Klaas-Jan Stol}.} \bibinfo{year}{2017}\natexlab{}.
\newblock \showarticletitle{{Continuous software engineering: A roadmap and
  agenda}}.
\newblock \bibinfo{journal}{\emph{Journal of Systems and Software}}
  \bibinfo{volume}{123} (\bibinfo{date}{jan} \bibinfo{year}{2017}),
  \bibinfo{pages}{176--189}.
\newblock
\showISSN{01641212}
\urldef\tempurl%
\url{https://doi.org/10.1016/j.jss.2015.06.063}
\showDOI{\tempurl}


\bibitem[\protect\citeauthoryear{Fowler}{Fowler}{2006}]%
        {Fowler2006Blog}
\bibfield{author}{\bibinfo{person}{Martin Fowler}.}
  \bibinfo{year}{2006}\natexlab{}.
\newblock \bibinfo{title}{Continuous Integration}.
\newblock
  \bibinfo{howpublished}{\url{https://www.martinfowler.com/articles/continuousIntegration.html}}.
\newblock


\bibitem[\protect\citeauthoryear{Humble}{Humble}{2010}]%
        {Humble2010Blog}
\bibfield{author}{\bibinfo{person}{Jez Humble}.}
  \bibinfo{year}{2010}\natexlab{}.
\newblock \bibinfo{title}{Continuous Delivery vs Continuous Deployment}.
\newblock
  \bibinfo{howpublished}{\url{https://continuousdelivery.com/2010/08/continuous-delivery-vs-continuous-deployment/}}.
\newblock


\bibitem[\protect\citeauthoryear{Humble and Farley}{Humble and Farley}{2010}]%
        {Humble2010}
\bibfield{author}{\bibinfo{person}{Jez Humble} {and} \bibinfo{person}{David
  Farley}.} \bibinfo{year}{2010}\natexlab{}.
\newblock \bibinfo{booktitle}{\emph{Continuous Delivery: Reliable Software
  Releases through Build, Test, and Deployment Automation}
  (\bibinfo{edition}{1st} ed.)}.
\newblock \bibinfo{publisher}{Addison-Wesley Professional},
  \bibinfo{address}{Boston, MA}.
\newblock
\showISBNx{0321601912}


\bibitem[\protect\citeauthoryear{Jacksman}{Jacksman}{2020}]%
        {Jacksman2020}
\bibfield{author}{\bibinfo{person}{Marie Jacksman}.}
  \bibinfo{year}{2020}\natexlab{}.
\newblock \bibinfo{title}{Continuous Integration Tools for Mobile vs Web.
  What’s the Difference?}
\newblock
  \bibinfo{howpublished}{\url{https://code-maze.com/ci-tools-for-mobile-apps/}}.
\newblock


\bibitem[\protect\citeauthoryear{Klepper, Krusche, Peters, Bruegge, and
  Alperowitz}{Klepper et~al\mbox{.}}{2015}]%
        {Klepper2015}
\bibfield{author}{\bibinfo{person}{Sebastian Klepper}, \bibinfo{person}{Stephan
  Krusche}, \bibinfo{person}{Sebastian Peters}, \bibinfo{person}{Bernd
  Bruegge}, {and} \bibinfo{person}{Lukas Alperowitz}.}
  \bibinfo{year}{2015}\natexlab{}.
\newblock \showarticletitle{{Introducing Continuous Delivery of Mobile Apps in
  a Corporate Environment: A Case Study}}. In \bibinfo{booktitle}{\emph{2015
  IEEE/ACM 2nd International Workshop on Rapid Continuous Software
  Engineering}}. \bibinfo{publisher}{IEEE}, \bibinfo{address}{Florence, Italy},
  \bibinfo{pages}{5--11}.
\newblock
\showISBNx{978-1-4673-7067-7}
\urldef\tempurl%
\url{https://doi.org/10.1109/RCoSE.2015.9}
\showDOI{\tempurl}


\bibitem[\protect\citeauthoryear{Leszko}{Leszko}{2017}]%
        {Leszko2017CdwD}
\bibfield{author}{\bibinfo{person}{Rafał Leszko}.}
  \bibinfo{year}{2017}\natexlab{}.
\newblock \bibinfo{booktitle}{\emph{Continuous delivery with Docker and Jenkins
  : delivering software at scale}}.
\newblock \bibinfo{publisher}{Packt Publishing}, \bibinfo{address}{Birmingham,
  UK}.
\newblock
\showISBNx{9781787126145}


\bibitem[\protect\citeauthoryear{Paul Duvall~and and Glover}{Paul Duvall~and
  and Glover}{2007}]%
        {Duvall2007}
\bibfield{author}{\bibinfo{person}{Steve~Matyas Paul Duvall~and} {and}
  \bibinfo{person}{Andrew Glover}.} \bibinfo{year}{2007}\natexlab{}.
\newblock \bibinfo{booktitle}{\emph{Continuous Integration: Improving Software
  Quality and Reducing Risk (Addison-Wesley Signature Series (Fowler))}}.
\newblock \bibinfo{publisher}{Addison-Wesley Professional},
  \bibinfo{address}{Boston, MA}.
\newblock


\bibitem[\protect\citeauthoryear{Shahin, {Ali Babar}, and Zhu}{Shahin
  et~al\mbox{.}}{2017}]%
        {Shahin2017a}
\bibfield{author}{\bibinfo{person}{Mojtaba Shahin}, \bibinfo{person}{Muhammad
  {Ali Babar}}, {and} \bibinfo{person}{Liming Zhu}.}
  \bibinfo{year}{2017}\natexlab{}.
\newblock \showarticletitle{{Continuous Integration, Delivery and Deployment: A
  Systematic Review on Approaches, Tools, Challenges and Practices}}.
\newblock \bibinfo{journal}{\emph{IEEE Access}}  \bibinfo{volume}{5}
  (\bibinfo{year}{2017}), \bibinfo{pages}{3909--3943}.
\newblock
\showISBNx{2169-3536 VO - 5}
\showISSN{2169-3536}
\urldef\tempurl%
\url{https://doi.org/10.1109/ACCESS.2017.2685629}
\showDOI{\tempurl}
\showeprint[arxiv]{1703.07019}


\bibitem[\protect\citeauthoryear{{Stack Overflow}}{{Stack Overflow}}{2016}]%
        {StackOverflowDeveloperSurvey2016}
\bibfield{author}{\bibinfo{person}{{Stack Overflow}}.}
  \bibinfo{year}{2016}\natexlab{}.
\newblock \bibinfo{title}{Developer Survey Results 2016}.
\newblock
  \bibinfo{howpublished}{\url{https://insights.stackoverflow.com/survey/2016}}.
\newblock


\bibitem[\protect\citeauthoryear{{Stack Overflow}}{{Stack Overflow}}{2018}]%
        {StackOverflowDeveloperSurvey2018}
\bibfield{author}{\bibinfo{person}{{Stack Overflow}}.}
  \bibinfo{year}{2018}\natexlab{}.
\newblock \bibinfo{title}{Developer Survey Results 2018}.
\newblock
  \bibinfo{howpublished}{\url{https://insights.stackoverflow.com/survey/2018}}.
\newblock


\bibitem[\protect\citeauthoryear{{Weber}, {Nepal}, and {Zhu}}{{Weber}
  et~al\mbox{.}}{2016}]%
        {Weber2016}
\bibfield{author}{\bibinfo{person}{I. {Weber}}, \bibinfo{person}{S. {Nepal}},
  {and} \bibinfo{person}{L. {Zhu}}.} \bibinfo{year}{2016}\natexlab{}.
\newblock \showarticletitle{Developing Dependable and Secure Cloud
  Applications}.
\newblock \bibinfo{journal}{\emph{IEEE Internet Computing}}
  \bibinfo{volume}{20}, \bibinfo{number}{3} (\bibinfo{year}{2016}),
  \bibinfo{pages}{74--79}.
\newblock
\urldef\tempurl%
\url{https://doi.org/10.1109/MIC.2016.67}
\showDOI{\tempurl}


\bibitem[\protect\citeauthoryear{Zhao, Serebrenik, Zhou, Filkov, and
  Vasilescu}{Zhao et~al\mbox{.}}{2017}]%
        {Zhao2017}
\bibfield{author}{\bibinfo{person}{Yangyang Zhao}, \bibinfo{person}{Alexander
  Serebrenik}, \bibinfo{person}{Yuming Zhou}, \bibinfo{person}{Vladimir
  Filkov}, {and} \bibinfo{person}{Bogdan Vasilescu}.}
  \bibinfo{year}{2017}\natexlab{}.
\newblock \showarticletitle{{The impact of continuous integration on other
  software development practices: A large-scale empirical study}}. In
  \bibinfo{booktitle}{\emph{2017 32nd IEEE/ACM International Conference on
  Automated Software Engineering (ASE)}}. \bibinfo{publisher}{IEEE},
  \bibinfo{address}{Urbana, IL, USA}, \bibinfo{pages}{60--71}.
\newblock
\showISBNx{978-1-5386-2684-9}
\urldef\tempurl%
\url{https://doi.org/10.1109/ASE.2017.8115619}
\showDOI{\tempurl}


\end{thebibliography}

\end{document}